# The Quantum Dilemma

Joseph J. Bisognano

Past Director of the Wisconsin Synchrotron Radiation Center and
Professor Emeritus of Engineering Physics at the University of Wisconsin
jjbisognano@wisc.edu

November 12, 2026

## Abstract

Physics has long lived with a split personality that desires determinism for measured systems while demanding that experimenters decide what to measure on a whim. Intriguingly, such a free will assumption for experimenters has thwarted many attempts to provide a satisfactory explanation of how quantum probabilities evolve to clear-cut measurements. An overview of this quantum measurement problem is presented without equations, and the lesson is drawn that denial of experimenters' free will may be the only workable solution. If the free will assumption is rejected, then a door is open that may ultimately reconcile quantum mechanics with the definiteness of individual experiments. A holistic view is offered of a self-consistent, superdetermined spacetime block of events. The goal of future research is to develop the underlying principle that generates this Escher-like global pattern.

Submitted to a Special Issue of the International Journal of Modern Physics A (IJMPA)

# Introduction

Paradoxically, it may be time to accept a more complete determinism to make full sense of the probabilistic description of quantum mechanics. While there are no contradictions, no inconsistencies with today's interpretation of quantum mechanics and experiments, the measurement act, which yields a specific response, is not described. A natural question to ask is whether there is some larger theory that would show how an individual experimental result is generated. All attempts to date to reconcile this quantum incompleteness have failed. Moreover, the works of Bell, Kochen, Specker, and Conway [1], [2], [3] provide severe constraints. For example, to be consistent with quantum predictions and a form of locality, the free will exhibited by experimenters in choosing what to measure requires that the measured system retains the freedom to decide what result to give. Physics has long lived with a tension that desires determinism for measured systems while experimenters decide on a whim. But this situation appears to be impossible. Rather than accepting that the observed system has a perplexing autonomy, it may be time to accept a form of complete determinism of action for both observer and observed to make full sense of quantum mechanics. Since quantum mechanical predictions only apply to ensemble averages, determinism for individual events is not forbidden, only uncomfortable. To quote Sherlock Holmes [4], "...when you have eliminated all which is impossible, then whatever remains, however improbable, must be the truth."

# The Quantum Dilemma

Quantum mechanics speaks to probabilities, where results of some measurements cannot be determined in advance; several outcomes are possible. Yet, at some point specific things happen. The equations of quantum mechanics only describe the evolution of the probabilities; the measurement act, which yields a definite result, is not described. One wonders whether there is some larger theory that would describe how a definite result is decided.

In and of itself, this situation might not be so conceptually troubling, since one might imagine that there is some random noise process that quantum mechanics describes in some average sense. This noise could be really random or pseudorandom (i.e., deterministic, but at a much finer level of resolution that we cannot control or measure today). Or, as described in various elementary quantum texts, the measurement process perturbs the system in an unknowable way, so that after an experiment yielding a definite result, other parameters that might be measured become uncertain, stirred up in the measurement process. Only some probability distribution of these parameters can be surmised.

Unfortunately, underlying this simple explanation there is inconsistency. It seems to say that the measured system really does have definite values, but we can't know them with certainty since they are always being jiggled around, either by some noise or by the measurement act. Way down deep, at any point in time, all the parameters have well-defined values. Such simplicity is, however, forbidden. There is a theorem, the Kochen-Specker Theorem [2], which says that this can't be the situation. In simple terms, this theorem says that there is no way for a system to have all possible experimental results determined in advance of the measurement, even at the microscopic level, and agree with what is observed in nature. The model of all parameters having definite values, but jiggled in an unknowable way, simply does not work. Very roughly, the most controversial assumption of this theorem is that values of compatible (commuting) variables are situationally independent (non-contextual)—a hint of free will. For example, noncontextuality implies that the value of the spin one angular momentum squared for in one direction is independent of which angular momentum squared in the orthogonal plane is measured. For entangled pairs with correlated variables, locality (finite correlation communication speed) seems to imply non-contextuality since a measurement of a component of angular momentum of one pair requires that measurement of the same component of angular momentum of a spacelike separated partner agree regardless of what other compatible measurements are done and in what order.

It appears that it's only through measurement that some of the observables become well defined. It's not that a particle "knows" at any point in time what value it will offer and that the uncertainty comes only through being jiggled from time to time or perturbed by an experimenter. There is something much more subtle. It seems that somehow the "negotiation" between system and measurer brings about the definite result. However strange, one might be able to accept that, but the finiteness of the speed of light provides even more uncomfortable corollaries.

Bell [1] considers a system that is broken into two subsystems that are moved apart in space. They are created (through some fundamental conservation principle) correlated or entangled so that if one subsystem (#1) has a property (usually quantum spin) measured with a particular value, the other subsystem (#2) has the comparable property measured with the negative of that value, summing to zero. This can be done in the laboratory and has been proven in many experiments. It turns out, according to Bell's proof, that having all possible experimental results determined in advance independently for each subsystem (hidden variables) can't describe what's observed in nature. A local hidden variable theory cannot match the predictions of quantum mechanics.

This is very similar to the Kochen-Specker result but offers a next level of discomfort that is hard to reconcile with the negotiation loophole. Consider experiments that are performed on #1 and #2 at times so that they are independent (that is, they cannot communicate because the speed of light speed is finite, so-called space-like separated). If we accept the negotiation idea, then experimenter #1 interacts with system #1, to yield a value of a property he/she chooses to measure. Experimenter #2, if he/she happens to measure the same property in a negotiation with system #2, must get the negative of that value. Yet experimenter #2 can't know the results of the negotiation underlying Experiment # 1, because there's no time to communicate (locality) and there can't be hidden variables predetermining the two negotiations. A requirement of the proof is that the particle must respond correctly to all experimental configurations that could have been set by the observer; there is again a component of free will.

Conway and Kochen [3] invoke the Kochen-Specker Theorem to get to a similar, possibly more troubling conclusion in their Free Will Theorem. It says explicitly that if an experimenter has the "free will" to pick an experiment to perform, then the system being measured also has the "free will" to decide what result to give.

To summarize: We have a perplexing situation: a) the measurement negotiation between system and experimenter seems to create the observable, the datum or the number, which cannot in general exist in advance, but b) the freely chosen measurements of spatially isolated (space-like separated) subsystems must be consistent, even when they cannot communicate. This apparently requires that somehow the measurement of a parameter of one subsystem forces the associated second subsystem to have completed a negotiation with a yet to be performed local experiment to set the comparable parameter. Correlations and negotiations look like they can be enforced faster than the speed of light.

## Is There a Problem?

Of course, one can argue, "so what." The Copenhagen interpretation of quantum mechanics lays out a scheme that has successfully predicted the answers to all the questions it poses. There are some questions that cannot be answered (like predicting the result of a single experiment), but those are declared to be unknowable in an absolute sense. What's wrong with accepting this and moving on? Also, although definiteness or correlation enforcement seems to propagate faster than the speed of light, information (communication) does not—relativity is unscathed. For the standard case of two subsystems in a zero angular momentum state, if one measures that one subsystem is spin up, one knows the other is spin down. Since it appears impossible to use this for communicating information, there is no obvious conflict with relativity.

The bottom line is that there are no contradictions, no inconsistencies with today's interpretation of quantum mechanics. Just something "spooky," as Einstein said. It may hint at something more, but the answer won't be the restoration of a deterministic world of physical systems as long as experimenters are assumed to have the freedom to pick any experiment they want to do. Here, possibly, is the one escape clause. After all, the past is determined and is consistent with quantum mechanics. We like to believe that the future has free choices, has free will. If we're willing to give that up, saying that the future is as determined as the past, then the impression of a paradox disappears. The randomness of a quantum measurement vanishes, but its predictions might be retained.

## Interpretations and the Measurement Problem

This fundamental tension between unitary evolution of probabilities and well-defined measurements remains unresolved despite numerous interpretations. One cannot even define the boundary between unitary evolution and irreversible measurements. Many interpretations attempt to reconcile measurement and state evolution. So far, all seems to be incomplete. For example,

- **Copenhagen**: The details of the measurement mechanism for the state collapse to an eigenstate are not explained. What determines whether an action is unitary or irreversible (nonunitary)?
- **Hidden Variables**: Conflicts with locality, experimental independence, and free will, as described by Bell and Kochen-Specker theorems, for example.
- **Many Worlds**: Ambiguous on probabilities and generates unlimited, weakly defined branching, including the many embodiments of the experimenter.
- **Bohmian mechanics:** Still suffers from nonlocality and requires special initial distributions.
- **Consistent Histories**: Decoherence does not generate a real collapse, just noninterfering probabilities; effectively Many Worlds.
- **Collapse Models**: Struggle with energy conservation and experimental constraints.
- **Transactional, Relational, and Informational (e.g., qBism) interpretations**: Often just sidesteps measurement definiteness and/or implicitly assumed hidden variables. The wavefunction is about knowledge and probabilities, but of what. Isn't this underlying system just implicitly hidden variables which must face Bell's theorem? One still needs a mechanism to generate the result of a specific roll of the dice.

## The Core Issues

Quantum mechanics functions probabilistically, yet experimental results are definite. Bell's and Kochen-Specker's theorems rule out classical hidden variable theories unless some locality or experimenter's free will is sacrificed. Alternative approaches seem to dance around this problem (the vast variety of "interpretations") or add theoretical extensions unsupported by experiments. Importantly, quantum mechanics and general relativity remain at odds because of this inherent nonlocality of quantum mechanics

Any full resolution must explain:

- How a measurement yields a definite result.
- Whether correlations of entangled states imply true nonlocality or a deeper pre-established structure, possibly violating an experimenter's free will.
- How to interface with General Relativity.

## Possible Future Directions

Here we list a few possibilities for future directions.

1. **Nonunitary Mechanisms**: Introduce self-consistent nonunitary processes to explain measurement definiteness. So far, this approach has not been successful. Environmental decoherence moves in the right direction, but leaves the final dice roll unexplained.
2. **Negotiation**: Define the mechanism of negotiation between a system and an experimenter. Investigate whether measurement as a negotiation between system and observer can be made to work for an entangled pair. Some nonlocality or interaction from the future seems to be necessary. Even if one gives up locality, it is not clear why distant entangled pairs would preferentially communicate.

3. **Discrete or emergent spacetime**: Modify the notion of spacetime to sidestep nonlocality of quantum mechanics. Quantum mechanics is fundamental, spacetime isn't. For example, spacetime could emerge from an S-matrix point of view with reality happening in momentum space (or alternatively, a holographic principle). Measurement is a nonunitary action on quantum Hilbert space outside of spacetime. Spacetime *emerges* with some "spooky" action at a distance that isn't fundamental.
4. **Superdeterminism and/or the Block Universe** Superdeterminism posits that all measurement settings and preparations are linked by self-consistency. Outcomes are predetermined, circumventing Bell's theorem by denying free will. A block universe perspective suggests a fully determined past and future consistent with quantum mechanics without communication but self-consistency as a spacetime painting or puzzle. There is a crystallization that results in a holistic spacetime consistent with quantum theory. However, this raises the question: why was the universe structured to obey quantum probabilities and interference patterns? What drives the process and determines the passage of time?

## Is Superdeterminism the Way?

Since the creation of relativity and quantum mechanics (and refinements such as quantum field theory) in the early part of the twentieth century, the stage for physics has remained static. There have certainly been important and powerful advancements. But relativistic spacetime, noncommuting measurements, and Hilbert spaces remain the platform. Nothing like the transition from classical physics has happened. To move forward and provide a satisfactory answer to the measurement problem, we must create a new stage that allows certainty for individual experiments while maintaining consistency with quantum predictions for ensemble averages. We have argued that this platform can be realized if we give up the assumption of the form of free will that physics continues to demand for the experimenter. Bell realized that such "superdeterminism" [5] provided the loophole, saying "There is a way to escape the inference of superluminal speeds and spooky action at a distance. But it involves absolute determinism in the universe, the complete absence of free will." More recently, 't Hooft and Hossenfelder have championed this idea [6],[7]. But Bell also saw this solution as improbable [1], arguing that in a situation, for example, where the experimental choices are made by a deterministic random number generator, "…it is unlikely for the hidden variable to be sensitive to all of the same small influences that the random number generator was." However, after decades of work, small progress has been made to reconcile the measurement problem at its most fundamental level. The consistent histories approach [8], together with decoherence phenomena, generates a form of quasi-certainty for macroscopic systems, but most basically remains probabilistic. The wavefunction of the universe never collapses to a well-defined result. The Many Worlds interpretation [9] generates an infinity of universes to answer the practical question of how a single measurement produces specific values. –And it's still not clear that the idea is fully well defined. We are indeed faced with the Holmesian realization that having "…eliminated all which is impossible, then whatever remains, however improbable, must be the truth."–And that truth is that the universe, for all space and time, is fully determined.

Replacing the free will assumption with a determinism assumption is only the beginning of this restructuring of the foundations of physical law. There are, no doubt, various paths that would be enabled by deciding that the free will assumption for experimenters is wrong. –And, of course, classical physics represents a deterministic framework that simply does not match reality. Historically, serious consideration of this approach has faced stymieing skepticism in the apparent conspiracy of initial conditions of the universe necessary to achieve quantum results without violating locality.

This alternative perspective substitutes global spacetime self-consistency. Consider a spacetime populated with a network of events that are self consistently laid out, as in a tessellated Escher print. Averaged over space-time, the events are consistent with quantum mechanics. Averaged over many individual results, experimental outcomes appear only probabilistically related to the preparation of the systems being measured. Averaged over many individual events, experimental choices appear uncorrelated with the preparation of the systems being measured—at some level experimenters show free will. Yet, for each individual experimental event, there is underlying certainty. Such a network must be possible, since at any fixed time, history is described by such a well-defined net fully consistent with quantum mechanics, and we would be simply extending this net to the full space-time "plane" rather than the "half-plane" of earlier times. One might imagine metaphorically that at an "instant of creation," this spacetime net is laid out as a whole, with the infinity of choices being made in a single step to create a particular universe, always being consistent with the rules of quantum mechanics. It is this singular "act of choice" that is experienced by an individual as free will in the time coordinate. What is perceived as free will is the fact that of the infinity of possible nets, the universe is just one specific representation.

## Conclusion

Interpretations of quantum mechanics typically do not lead to new experimental predictions and only offer some additional comfort in accepting the Copenhagen predictions of quantum physics. There remain no cogent answers to the question of how an individual experimental result is generated. Many mathematical constraints exist that have thwarted all attempts to satisfactorily answer this question. Accepting that the experimenter's free will assumption is wrong offers a powerful foundation to move forward. This assumption might be replaced, for example, by a self-consistent spacetime net that provides definiteness for individual experiments while exhibiting appropriate correlations and independence when averaged over many similar events. Free will is exhibited by an experimenter's decisions being uncorrelated with the history of the measured system in an ensemble average. With all of spacetime truly a unified whole, existing outside of the flow of time, determinism and quasi free will coexist. This approach might ultimately yield areas to test the completeness of quantum mechanics, asking questions that quantum mechanics simply does not pose.

Although absolutely proving impossibility may be beyond the realm of physical argument, we conjecture that, after decades of work and a growing body of no-go theorems, solving the quantum measurement problem is likely impossible without dropping the assumption of an experimenter's free will. Many have realized that superdeterminism is the loophole, but discomfort with initial condition conspiracies forces these ideas out of the mainstream of research. The core message of this short essay is for the physics community to grasp this idea wholeheartedly and see where it leads. We offer one example, the block universe. But this is only given as an example, and other ideas, current and future, may ultimately be the way forward. The main point is to urge a serious investigation of where the denial of an experimenter's free will can take us. This improbable solution may prove to be the only possible solution.

## Note

Swapan Chattopadhyay and I have been corresponding and interacting on these issues regularly over the past few years. The give and take of these discussions have provided key ideas for my pursuit of the measurement problem in quantum mechanics. Swapan invited me to contribute to this Special Issue in a form without equations for a general audience.